# Prediction of void evolution in sheet bending based on statistically representative microstructural data for the Gurson-Tvergaard-Needleman model

A. Schowtjak[1*], C. Kusche[2], R. Meya[1], S. Korte-Kerzel[2], T. Al-Samman[2], A. E. Tekkaya[1] and T. Clausmeyer[1]

[1]Institute of Forming Technology and Lightweight Components, TU Dortmund, Germany
[2]Institute of Physical Metallurgy and Metal Physics, RWTH Aachen, Germany
*Corresponding Author (Alexander.Schowtjak@iul.tu-dortmund.de)

## Abstract

Ductile damage in sheet steels is caused by voids. It is crucial for product design to predict the distribution of voids in bent components. Since the void volume fraction is a state variable in the Gurson-Tvergaard-Needleman (GTN) model, it is applied to predict the evolution of voids in bending. Material parameters are identified based on force-displacement curves of a dual phase steel and also through statistical microstructural information obtained from panoramic scanning-electron microscopy images. The void volume fraction and particular void populations of GTN-model are determined with a recently proposed scheme, which involves machine learning algorithms.

## 1. Introduction

The determination of the material and model parameters is of great importance for accurate prediction, since it directly influences the elasto-plastic material behavior as well as the damage evolution. In this work, the dual phase steel DP 800 is investigated. The void volume fraction is determined experimentally by scanning electron microscopy (SEM). Here, two different strategies, i.e. a macroscopic and a microscopic approach, are introduced, analyzed and finally validated in terms of an air bending experiment.

## 2. Gurson porous plasticity

The Gurson-Tveergard-Needleman (GTN) model is a micromechanics-based damage model [1], where $\Phi^p = [\sigma_{eq}/\sigma_y] + 2\,q_1 f\,\cosh[3q_2\,\sigma_h/(2\,\sigma_y)] - [1 + q_3 f^2]$ is the yield function, with $q_1, q_2$ and $q_3 = q_1^2$ representing material parameters and $f$ denoting the effective void volume fraction. Since no coalescence was observed for the investigated levels of deformation, the terms for coalescence are neglected and consequently the evolution relation for the void volume fraction is given by

$$\dot{f} = \underbrace{\frac{f_N}{s_N\sqrt{2\pi}}\exp\left[-\frac{1}{2}\left(\frac{\varepsilon_{eq}^p - \varepsilon_N^p}{s_N}\right)^2\right]\dot{\varepsilon}_{eq}^p}_{\dot{f}^n} + \underbrace{[1-f]\,\mathrm{tr}(\dot{\boldsymbol{\varepsilon}}^p)}_{\dot{f}_{hyd}^g} + \underbrace{k_w\,f\,\frac{w(\mathrm{dev}(\boldsymbol{\sigma}))}{\sigma_{eq}}\,\mathrm{dev}(\boldsymbol{\sigma}){:}\boldsymbol{\varepsilon}^p}_{\dot{f}_{shr}^g},$$

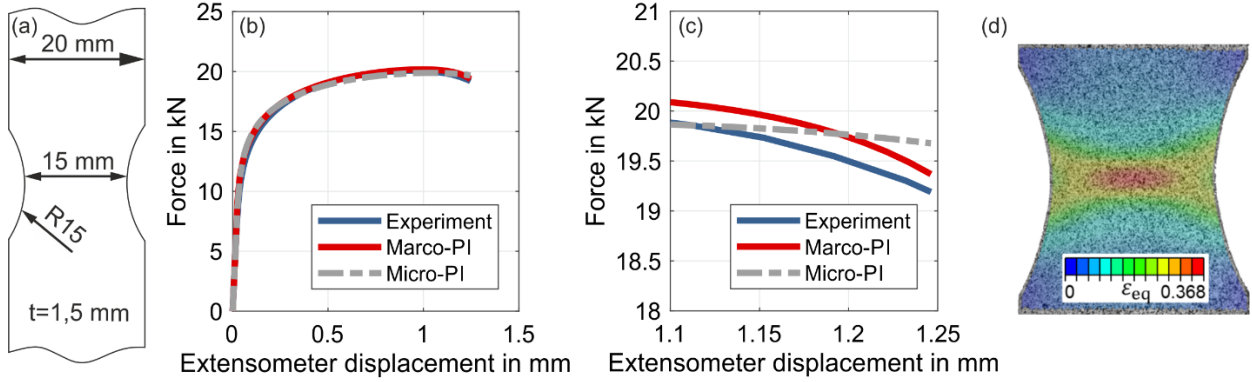

*Figure 1: Specimen for notched tensile test (a), load displacement curve for the experimental and the simulation for the whole test (b) and (c) as well as DIC data right before failure (d).*

with $w(\text{dev}(\boldsymbol{\sigma})) = 1 - [27 J_3/2 \sigma_{eq}^3]^2$, where $J_3$ is the third deviatoric stress invariant. It is composed of a growth part for hydrostatic and shear stresses, i.e. $\dot{f}_{\text{hyd}}^g$ and $\dot{f}_{\text{shr}}^g$, respectively, as well as a nucleation-related term $\dot{f}^n$. The latter one is a Gaussian curve with $\varepsilon_N^p$ being the mean value of plastic strain at maximal nucleation [2]. The standard deviation is denoted by $s_N$ and $f_N$ scales the nucleated volume fraction.

## 3. Parameter identification

Two different strategies for the parameter identification (PI) are investigated, i.e. a macroscopic and a microscopic approach. The parameter identification is performed inversely using a notched tensile test to induce necking and an inhomogeneous stress state, see Fig. 1 (a). In order to avoid local minima in the optimization process, a simplex algorithm with various initial parameter sets as well as evolutionary algorithms are used. The specimen was pulled up to fracture with a nominal strain rate of $0.01/s$. Digital image correlation (DIC) data is used in order to measure the displacement field within the notched area, as depicted in Fig. 1 (d). The specimen's elongation is computed based on this data in the region outside of the notch, where the displacement is homogeneous. Post-mortem REM-analysis was performed for the failed specimen.

The $q_1 = 1.5$ and $q_2 = 1.0$ are chosen as proposed by Tvergaard [3]. Since shear is not a dominant stress state in a notched tensile test, the associated variable is not identified here and $k_w$ is defined as 1.0. The nucleation-related parameters, i.e. $f_N$, $s_N$ and $\varepsilon_N^p$, are to be identified. Additionally, the parameters associated with plasticity are identified. Hardening according to Swift, i.e. $\sigma_y = A[\varepsilon_0 + \varepsilon_{eq}^p]^n$, is assumed. The Young's modulus $E$ as well as the Poisson's ratio $\nu$ are chosen as 210 GPa and 0.3, respectively.

For the macroscopic approach, all parameters are identified inversely solely based on the experimental load-displacement curve. The numerical results, as depicted in Figure 1 (a),

*Table 1: Parameters identified for the macroscopic (set 1) and microscopic (set 2) approaches.*

|  | $A$ in MPa | $\varepsilon_{eq}^p$ | $n$ | $f_N$ | $s_N$ | $\varepsilon_N^p$ |
|---|---|---|---|---|---|---|
| Set 1 | 1336.6 | $4.044 \cdot 10^{-4}$ | 0.1610 | 0.075 | 0.2933 | 0.1854 |
| Set 2 | 1242.8 | $1.031 \cdot 10^{-5}$ | 0.1390 | 0.099 | 0.1172 | 0.6168 |

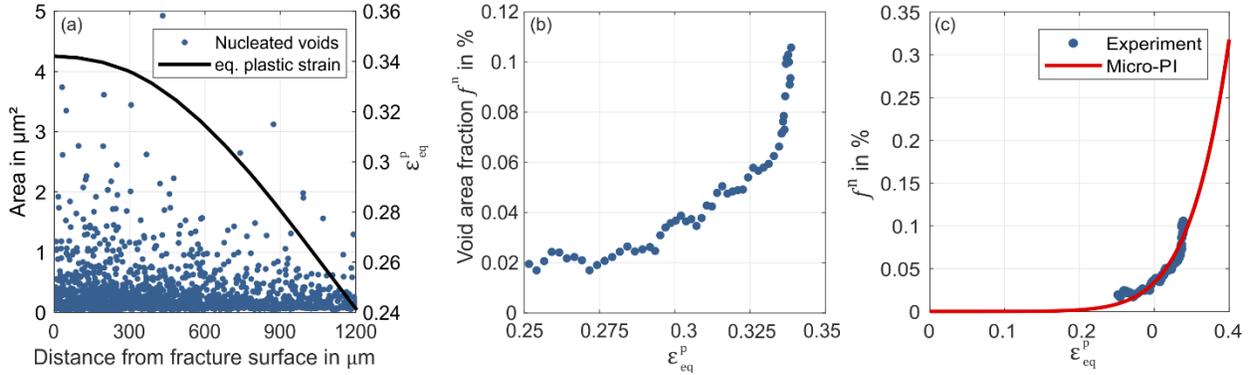

*Figure 2: Experimental data for nucleated voids and strain measurement (a), evolution of nucleated voids due to nucleation (b) and numerical results for the microscopic approach (c).*

are in good agreement with the experimental data. The identified parameter set is shown in Table 1.

To resolve the microscopic voids, high resolutions only achievable by scanning electron microscopy are necessary. In order to account for statistical representativeness, the problem is addressed by employing a panoramic imaging approach that divides the observed area into individual, overlapping frames. Those frames are joined together with an image correlating algorithm, resulting in a field of view of 335 x 1200 µm. Specifically developed image processing is applied to measure the area of single voids in separate frames of 8x8 µm [4]. The void area fraction is calculated using overlapping, moving bins. On the basis of the mean void size in the previous bin, the threshold for newly nucleated voids is calculated. In these investigations, 1685 voids are classified as nucleated, as depicted in Fig. 2 (a).

Since the rate of deformation is unknown, the nucleation-related parameters cannot be determined by solely considering the microscopic data, i.e. $\dot{f}^n$ and $\dot{\varepsilon}^p_{eq}$. Hence, the parameters are identified in an inverse manner by performing FE-simulations of a single unit cell under tension. The effect of the stress state on the void evolution is neglected here. To this end, the void area fraction is calculated and mapped to the equivalent plastic strain distribution using the DIC data, see Fig 2 (b). However, it is assumed that the strain distribution throughout the thickness of the specimen is homogeneous. Based on this data, the nucleation-based parameters are identified. It is assumed that the void volume and area fraction are equal. Subsequently, the plasticity related variables are determined inversely based on the load displacement curve of the notched tensile test, see Table 1. While the evolution of the nucleated voids for the microscopic approach is in good agreement with the experiment, as depicted in Fig. 2 (c), the void volume fraction predicted with the macroscopic approach is 2 orders of magnitude higher. The macroscopic behavior, however, i.e. the load displacement data, is more accurate for the first approach. Especially the softening is accounted for more precisely, see Fig. 1 (c).

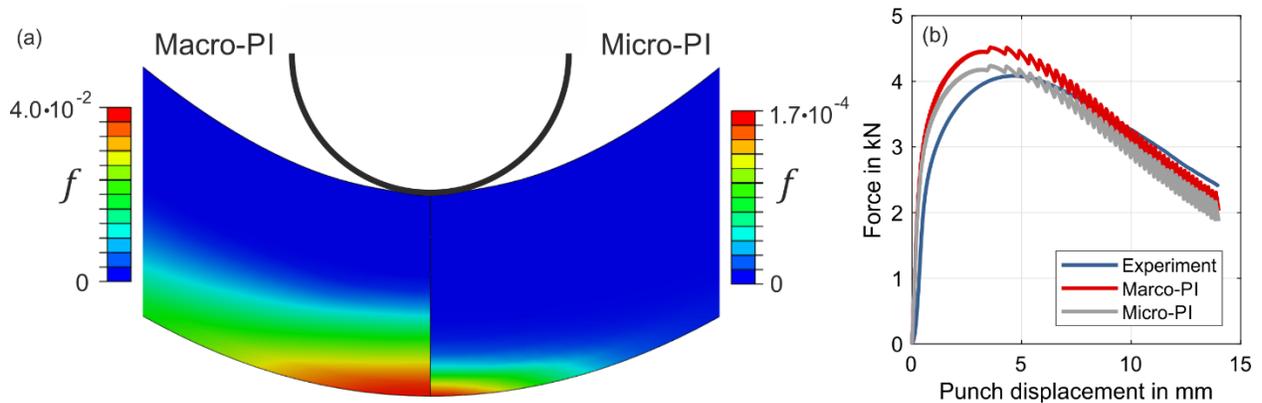

*Figure 3: Void volume fraction distribution in the bent component as well as the load displacement curve of the bending experiment (b).*

## 4. Bending

Air bending experiments (punch and die radii $r = 1$ mm, die width $w = 24$ mm with a punch speed of $v = 0.3$ mm/$s$). While the void volume fraction in the compressive zone is zero, both simulations show its maximum on the outer radius, as depicted in Fig. 3 (a). For the simulation with the parameters obtained with the microscopic approach, the predicted void volume fraction is more realistic, since it is in the order of magnitude of the experimental results of the tensile test. The void volume fraction obtained with the macroscopic identification approach is approximately 100 times higher. Both load-displacement curves, as shown in Fig. 3 (b), are in good agreement with the experiment. Since the void volume fraction for the simulation with the microscopic approach is rather low, there is no damage induced softening. The simulation for the macroscopic strategy shows a larger increase in force at the start of the loading as well as a larger decrease in force towards the end of the process due to mechanical softening.

## 4. Conclusion

The parameters obtained with the macroscopic approach lead to a better fit with the load-displacement curves, while the parameters identified with the microscopic scheme represent the experimentally measured void area fraction better. The here introduced version of the GTN-model correlates the damage-induced softening to the void volume fraction. In future works, this correlation will be investigated in more detail.

**Acknowledgements**. This work was funded by DFG SFB/TR 188 (projects A05, B02 and S01).